\documentclass[preprint,12pt,5p,twocolumn]{elsarticle}

\usepackage{amssymb}
\usepackage{amsthm}
\usepackage{xcolor}
\usepackage{color,soul} 
\newcommand{\tildetext}{\raisebox{0.5ex}{\texttildelow}}
\usepackage[switch]{lineno} 
\usepackage{hyperref}
\usepackage{aasmacros}

\journal{High Energy Density Physics}

\begin{document}

\begin{frontmatter}

\title{Pseudoatom Molecular Dynamics Plasma Microfields}


\author[utaustin,lanl]{J. R. White \corref{cor1}}
\ead{jacksonw@lanl.gov}
\author[lanl]{C. J. Fontes}
\author[lanl]{M. C. Zammit}
\author[utaustin,cuboulder,nso,Hale]{T. A. Gomez}
\author[lanl]{C. E. Starrett}

\affiliation[utaustin]{organization={Department of Astronomy, University of Texas at Austin},
            city={Austin TX},
            postcode={78712}, 
            country={USA}}

\affiliation[lanl]{organization={Los Alamos National Laboratory},
            city={Los Alamos NM},
            postcode={87545}, 
            country={USA}}

\affiliation[cuboulder]{organization={Astrophysical and Planetary Sciences, University of Colorado Boulder},
            city={Boulder CO},
            postcode={80309}, 
            country={USA}}
\affiliation[nso]{organization={National Solar Observatory, University of Colorado Boulder},
            city={Boulder CO},
            postcode={80309}, 
            country={USA}}
\affiliation[Hale]{organization={George Ellery Hale Fellow}}
\cortext[cor1]{Corresponding Author}

\begin{abstract}
Spectral line profiles are powerful diagnostic tools for both laboratory and astrophysical plasmas, as their shape is sensitive to the plasma environment. The low-frequency component of the electric microfield is an important input for analytic line broadening codes. In this paper we detail a new method of calculating plasma microfields using configuration-resolved pseudoatom molecular dynamics. This approach accounts for both quantum atomic structure and N-body effects, similar to density functional theory molecular dynamics, but with less computational cost. We present pseudoatom microfields at conditions relevant for recent laboratory experiments. Compared to established microfield codes we find moderate deviations at solid density conditions and strong agreement at lower plasma densities.

\end{abstract}


\end{frontmatter}

\section{Introduction}


Spectral lines describe the opacity distribution of bound atomic transitions in plasmas. Fitting line shape broadening profiles to data enables detailed characterizations of plasma temperatures, densities, and compositions. This technique is widely used in both laboratory experiments \citep{Schaeuble19,Nagayama19} and astrophysical observations \citep{Genest_Beaulieu2019,Blouin20}. Spectral line broadening also impacts Rosseland mean opacities \citep{Krief16,Gomez18_Rosseland} and radiation transport in high energy density (HED) regimes. 

The dominant broadening mechanism in many HED plasmas is Stark broadening; fluctuations in the charged particle distribution generate perturbing electric fields, which shift the energy levels of radiating atoms. These small-scale perturbing electric fields are called the microfield. The time-averaged and volume-averaged microfield is zero by definition, but the temporal and spatial variations in the microfield are determined by the plasma conditions. Because of their direct influence on line shape broadening \citep{Gomez2022}, accurate descriptions of the microfield are important HED diagnostic tools. 

Analytic spectral line shape models make a distinction between the `high-frequency' microfield generated by fast moving electrons and the `low-frequency' microfield generated by ions and ion-electron correlations, which evolve on slower timescales. Following Griem's formulation \citep{Griem1959}, the analytic fundamental line shape equation is given as
\begin{equation}
    I(\omega) = \int_0^\infty P(\epsilon) J(\epsilon,\omega) d\epsilon,
    \label{eq:fund_ls_eq}
\end{equation}
where $I(\omega)$ is the line strength at frequency $\omega$, $P(\epsilon)$ is the low-frequency microfield distribution, and $J(\epsilon,\omega)$ is the high-frequency broadened line shape for a radiating atom located in an external electric field of constant magnitude $\epsilon$. $P(\epsilon)$ is a simple probability distribution. This low-frequency microfield probability distribution is often referred to simply as `the microfield'. We will adopt the same convention throughout this work.

Beginning with the Holtsmark distribution \citep{Holtsmark19}, which considered a plasma of charged particles with uncorrelated positions, many microfield models have been developed. The most common forms come from (see the review of \citet{Demura09} for a more comprehensive list) molecular dynamics (MD) simulations \citep{Stambulchik07,HauRiege17,Calisti24}, Monte-Carlo (MC) methods \citep{Brush66,Potekhin2002}, and the adjustable parameter exponential approximation (APEX) \citep{Iglesias1985,Iglesias2000}.

In this paper we present a new method of calculating the electric microfield distribution for hot dense plasmas using a configuration-resolved pseudoatom molecular dynamics (PAMD) model. PAMD \citep{Starrett2013,Starrett14,Starrett15} has previously been used to calculate equations of state and transport properties of dense plasmas. It is similar to density functional theory molecular dynamics (DFT-MD), but is computationally more efficient because it populates the simulation with pseudoatoms, the structure of which does not need to be recalculated at each time step.

Pseudoatoms are uniquely applicable to the calculation of low-frequency microfields. In the pseudoatom model, the free electrons are treated as a spherically symmetric screening cloud around each nucleus, and ion-electron correlations are accounted for using linear response theory. The high-frequency contribution to the microfield is then naturally neglected and the time history of inter-pseudoatom forces automatically gives the low-frequency electric microfield.

To date, the pseudoatom electron densities have been calculated using a DFT-based, average atom model.  Since it is DFT-based, the pseudoatom states have Fermi-Dirac occupations, which are in general, fractional. We now introduce configuration-resolved pseudoatoms, meaning the pseudoatoms can have integer bound state occupations and therefore PAMD may be carried out with arbitrary electron configurations.

This paper is split up into 5 sections. In section \ref{sec:theory} we detail the theory behind configuration-resolved pseudoatom molecular dynamics. In section \ref{sec:results} we present plasma microfields calculated with PAMD, compare them against other established microfield techniques, and demonstrate their impact on line shape profiles. In section \ref{sec:discussion} we offer a brief discussion of our results. Finally in section \ref{sec:conclusion} we present our conclusions.

Atomic units are used unless otherwise stated where $\hbar = m_e = k_B = e = a_0 = 1$.

\section{Configuration-Resolved Pseudoatom Molecular Dynamics} \label{sec:theory}

PAMD models a plasma as a collection of pseudoatoms. A pseudoatom is a charge-neutral object comprised of a nucleus with charge $Z_{nuc}$ and a surrounding spherically symmetric electron distribution $n_e^{PA}(r)$, which includes both bound electrons (the ``ion'' electron density) and free electrons (the ``screening'' electron density). 
Hence, for a pseudoatom species of subscript $i$ the electron distribution
\begin{equation}
    n_{i,e}^{PA}(r) = n_{i,e}^{scr}(r) + n_{i,e}^{ion}(r)
\end{equation}
satisfies
\begin{equation}
    \int n_{i,e}^{PA}(r) d^3r = Z_{i,nuc}.
\end{equation}
In this work, a species is defined by its nuclear charge $Z_{i,nuc}$ and its electron configuration $c_i$. Configuration definitions are discussed in greater detail in section \ref{sec:configs}.

The effective pair potential between pseudoatoms of different species in a mixture is derived from the quantum Ornstein-Zernike equations \citep{Chihara84} in \citet{Starrett14},
\begin{equation}
    V_{ij}(k) = \frac{4\pi \bar Z_i \bar Z_j}{k^2} - \frac{C_{ie}(k)}{\beta}n_{j,e}^{scr}(k),
    \label{eq:potential_eq}
\end{equation}
where the potential is calculated in momentum space for pseudoatom species $i$ and $j$, ion-electron direct correlation function $C_{ie}$, inverse temperature $\beta$, and the integrated screening electron density $\bar Z$ of each species,
\begin{equation}
    \bar Z_i = \int n_{i,e}^{scr}(r) \; d^3r.
\end{equation}

Mixtures are constrained by the requirement that the chemical potentials of the electrons in each pseudoatom are equal,
\begin{equation}
    \mu_i = \mu,
    \label{eq:mu_equal}
\end{equation}
and by the specified average mass density $\rho$,
\begin{equation}
    \rho = \sum_i f_i \, \rho_i(\mu,T),
    \label{eq:density_req}
\end{equation}
where $f_i$ is the species number fraction, $\rho_i$ the mass density for each species, and $T$ is the plasma temperature.

The electronic structure for a given species in the plasma is calculated using DFT-based techniques as detailed in \citep{Starrett2013,Starrett14,Starrett15}; it is not recalculated between simulation time steps. Perturbations to the screening electron density are accounted for in the calculation of the internuclear potential through the ion-electron correlation function (Eq. \ref{eq:potential_eq}). Molecular dynamics simulations can therefore be carried out, with quantum electrons, at computational speeds that are orders of magnitude faster than DFT-MD. The PAMD computational cost is equivalent to a classical MD simulation with precomputed inter-atomic potentials. A typical PAMD run, with a few hundred pseudoatoms, can complete a few million timesteps per hour on a single compute node.

\subsection{Configuration Definition} \label{sec:configs}
We now discuss the configuration definition and our implementation of configuration-resolved pseudoatoms.
We define a bound electron configuration $c$,
\begin{equation}
    c \equiv \prod_s (n_s l_s m_s)^{q_s}
    \label{eq:config_eq}
\end{equation}
where $n_s$, $l_s$, and $m_s$ are respectively the principal, azimuthal, and magnetic quantum numbers for state $s$. $q_s$ denotes the occupation number. We note that a state $s$ is sometimes instead referred to as an `orbital'.

\begin{figure}
    \centering
    \includegraphics[width=\columnwidth]{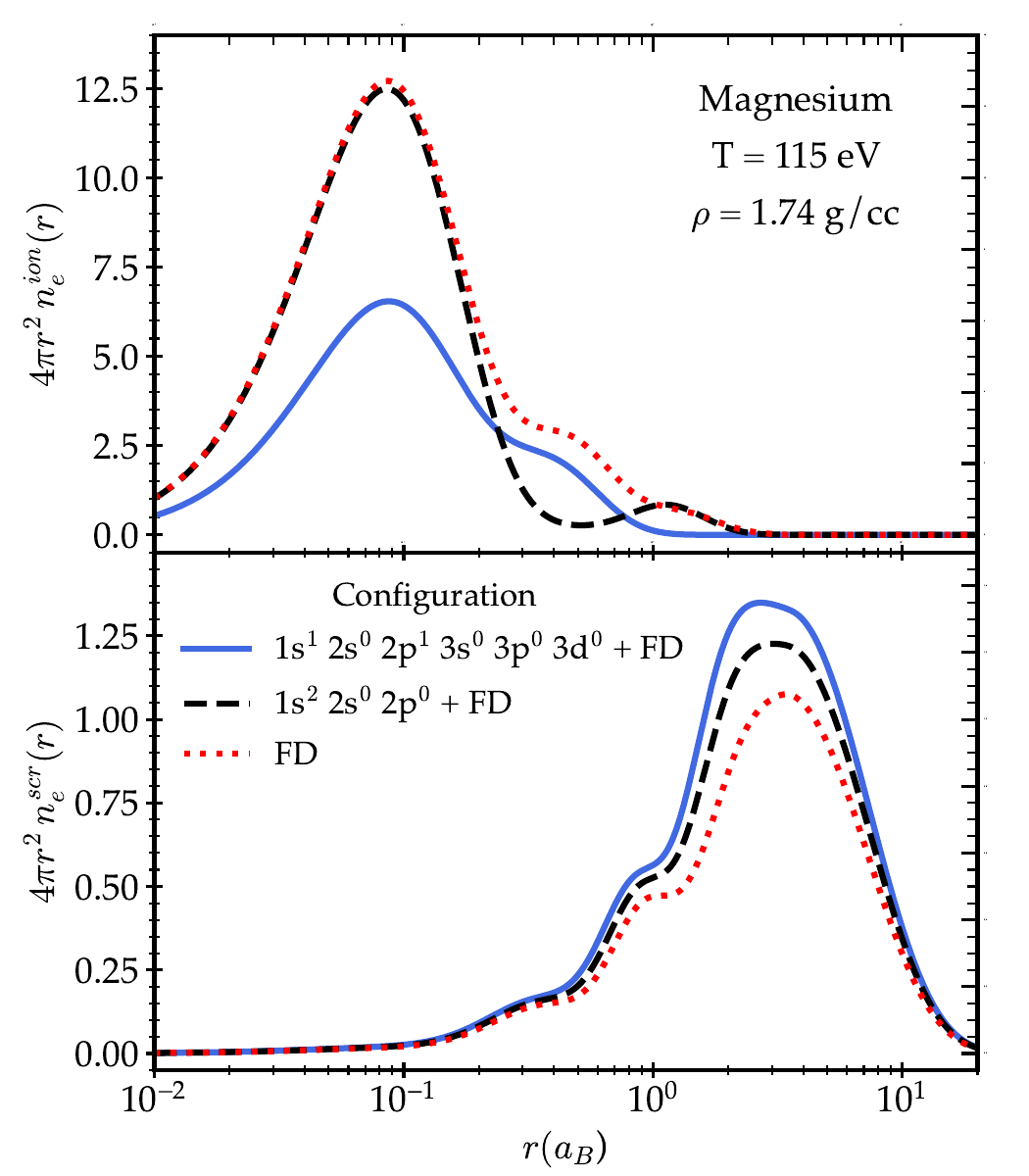}
    \caption{Electron densities for three different Mg configurations in a solid density Mg plasma.} 
    \label{fig:ne_Mg}
\end{figure}

The occupation numbers $q_s$ can be set in multiple ways. They are either specified by the user on input, calculated using Fermi-Dirac (FD) statistics, or taken as a mixture of the two. The Fermi factor $f$ for a state $s$ is given as
\begin{equation}
    f_s = \frac{1}{e^{(E_s - \mu)\beta}+1},
    \label{eq:FD_occupation_factor}
\end{equation}
using the energy of the state $E_s$, the chemical potential of the plasma $\mu$, and the inverse temperature $\beta$. The FD occupation is then simply
\begin{equation}
    q_s^{FD} = f_s \; g_s,
\end{equation}
where $g_s$ is the degeneracy of the state. A species defined by FD occupation numbers is equivalent to a pseudoatom as employed in \citet{Starrett15}. However, by specifying integer occupation numbers on input we can also consider pseudoatoms with specific electron configurations, which we refer to as `configuration-resolved' pseudoatoms. Configuration resolution is known to be advantageous over an average atom model, particularly in the calculation of opacities \citep{Hansen23}.

In figure \ref{fig:ne_Mg}, we show the bound and free electron density for three magnesium (Mg) configurations located in a solid-density Mg plasma. We use the same notation as in \citet{Starrett24}, where `+ FD' denotes that all higher order states (both above and below the continuum) are occupied using FD statistics. While the three configurations have the same $Z_{nuc}$ and similar $\bar Z$ values, the variations in bound electron configuration induce different structure in the electron densities.

\subsection{Valence Shell Treatment} \label{sec:valence_shell}


Transitioning across pressure ionization boundaries is challenging when defining atoms based on their bound electron configuration.
Discontinuities often arise because bound states and pressure ionized states are treated differently. For example, $\bar Z$ is often discontinuous across changes in density \citep{Starrett18}. 

In our configuration-resolved model, tightly bound states are occupied according to integer occupation numbers. Heavily pressure ionized states contribute to the screening electron density according to FD statistics. Suddenly transitioning between these two definitions, when a state is pressure ionized, causes sudden discontinuities.
This behavior is both unphysical and computationally unfavorable as it can lead to numerical problems when calculating plasma transport properties derived from configuration-resolved pseudoatoms.

We explore two treatments of the valence shell electrons that attempt to solve this problem. The first is the treatment described in the Excited State Model (ESM) \citep{Starrett24}, where the density of states (DOS) is filled according to integer occupations up to and including some $n_{max}$. Beyond $n_{max}$ the DOS is filled according to FD occupations regardless of the classification of states as `bound' or `pressure ionized'. The second is a `mixing' procedure, where a pressure broadening width is used to transition between integer occupations and FD occupations when a bound state approaches the pressure ionization limit. 

In the mixing procedure we calculate a continuum mixing probability \citep{Starrett2013}, 
\begin{equation}
    M(E_s) = \mathrm{erf}\Bigg(\frac{-2 \sqrt{\mathrm{ln2}}E_s}{\gamma}\Bigg),
    \label{eq:mixing_eq}
\end{equation}
where erf denotes the error function, $E_s$ is the energy of a bound state, and $\gamma$ is the inverse average relaxation time. 
Physically, $\gamma$ is equivalent to a broadening width. The mixing factor $M(E_s)$ can be thought of as the probability that the energy of state $s$ is pushed above the local continuum via pressure broadening, even though $E_s$ is below the local continuum on average. Currently, the broadening width is calculated as $\gamma = n_e^0 / \sigma_{dc}$, where $n_e^0$ is the free electron density in the limit that you move far away from a nucleus, and $\sigma_{dc}$ is the dc conductivity \citep{Starrett2013}. The mixing occupation factor for bound states is then taken to be
\begin{equation}
    q_s^{mix} = M(E_s)q_s^{input} + (1-M(E_s))q_s^{FD}.
\end{equation}
By gradually pushing states into the continuum as the density increases, this procedure helps smooth the transition in occupation number.

\begin{figure}
    \centering
    \includegraphics[width=\columnwidth]{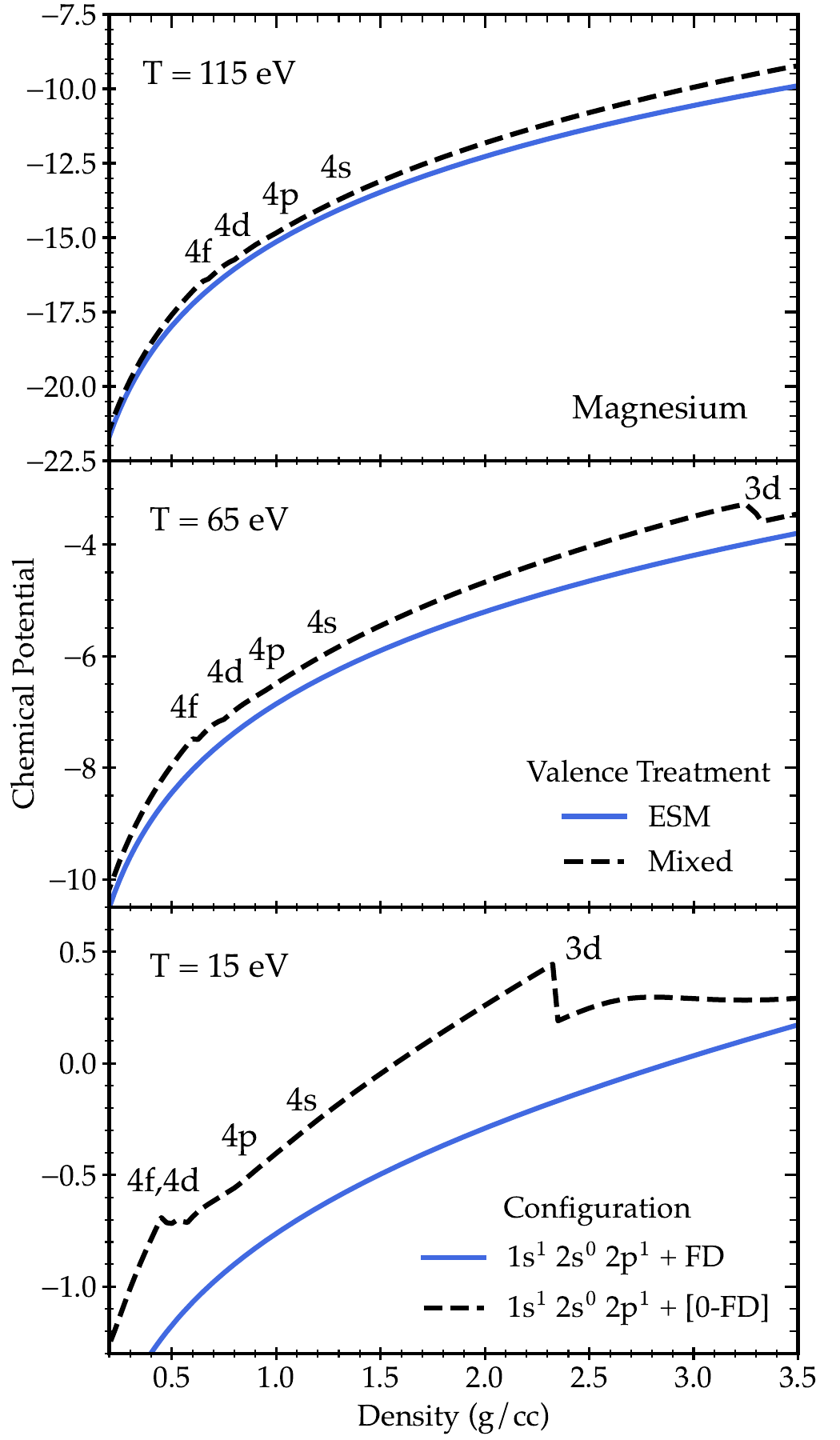}
    \caption{Magnesium chemical potential curves over a range of densities and three temperatures. The solid curves employ the ESM model of the valence shell while the dashed line uses the mixing approach, detailed in section \ref{sec:valence_shell}. Pressure ionization boundaries of different states are labeled on each plot.}
    \label{fig:MgTall_chem_pot}
\end{figure}

The ESM uses the notation `+ FD' to denote that higher order states in a configuration are occupied according to FD statistics. For the mixing model, we use a similar notation `+ [0-FD]' to denote that all higher order states are assumed to be empty when tightly bound and transition to FD occupations when pressure ionized. This notation is demonstrated in the configurations of figure \ref{fig:MgTall_chem_pot}.

The benefit to using these valence shell treatments can be seen in the calculation of chemical potential curves.
In a multi-species plasma, the chemical potential of each component species must increase monotonically with density to ensure that there is a unique solution for each pseudoatom (Eq. \ref{eq:mu_equal} and \ref{eq:density_req}). 
Multi-valued solutions for the effective density of an individual species can otherwise be found for a specified chemical potential. 
As the density of a plasma increases, more atomic states are pressure ionized and sudden jumps in the chemical potential curves can arise due to discontinuous occupations. Ensuring smoothly varying electron occupations helps prevent such jumps in the chemical potential curves.

Figure \ref{fig:MgTall_chem_pot} plots the chemical potential as a function of density for a Mg pseudoatom in a Mg plasma at temperatures of 10~eV, 50~eV, and 150~eV respectively. 
In each plot the chemical potential is calculated once using the ESM approach (with $n_{max}=2$) and once using the mixing approach. At some conditions, particularly at low temperatures, we find that a smooth transition can only be found with the ESM approach. Our mixing procedure is generally successful at preventing discontinuities at higher temperatures, but the relatively simple broadening width $\gamma$ is sometimes inadequate at lower temperatures.

In the ESM approach, our choice of $n_{max}$ is separate from any definition of the pressure ionization boundary. The appropriate value of $n_{max}$ is not always an obvious choice. In figure \ref{fig:MgTall_chem_pot}, for example, the model contains bound $n=3$ states despite our choice of $n_{max} = 2$. This is physically justified if the FD-occupied bound states are hybridized, whereby the states in question are bound but shared across multiple nuclei. 
For solid density Mg, comparison with laboratory experiments \citep{Ciricosta16} suggests that the valence shell is indeed hybridized and ought to be modeled with FD occupations, as discussed in \citet{Thelen24}. 
This assumption of orbital hybridization is also substantiated by the fact that the $n=3$ wavefunctions extend well beyond the Wigner-Seitz radius at these conditions. Determining strict limits for when a given choice of $n_{max}$ is valid is, however, challenging and we do not attempt to provide a robust answer here. 

We wish to emphasize that both solutions discussed for modelling valence shell electrons are crude approximations that attempt to correct for the largely unphysical distinction between loosely `bound' states and slightly `pressure ionized' states. In the future a more physical description of heavily perturbed electronic states, especially one that moves away from the convention where states are necessarily bound or pressure ionized, would be desirable.

\subsection{Microfield Calculation} \label{sec:microfield_results}

We refer to the pseudoatom at which we measure the microfield as the ``radiator'', while the pseudoatoms that surround the radiator are referred to as ``perturbers''. The molecular dynamics simulations presented in this paper are carried out by populating a periodic boundary-condition simulation cell with a single configuration-resolved radiator and many FD-occupation perturbers.

We define the net electric field felt by the radiator at time $t$ as the magnitude of the summed inter-atomic Coulomb forces, divided by the net charge of the radiator,
\begin{equation}
    \epsilon_i(t) = |\vec \epsilon_i(t)| = \bar Z_i^{-1} | \sum_j \frac{d V_{ij}}{dr} \hat r_{ij} |,
    \label{eq:microfield_equation}
\end{equation}
where $i$ denotes the radiator and the sum over $j$ denotes a sum over all perturbers. $V_{ij}$ was given in equation \ref{eq:potential_eq}. Physically, the first term in equation \ref{eq:potential_eq} accounts for the ion microfield, while the second term accounts for ion-electron correlations and serves as the screening mechanism. Together they give the `low-frequency' component of the microfield. 

In comparison to classical MD simulations \citep{Calisti11,HauRiege15}, this model does not resolve individual free electrons. Our treatment is, however, advantageous for a number of reasons.
First, we treat the free electrons quantum mechanically rather than classically. Second, we can take larger timesteps as we only need to resolve the ion trajectories, thus enabling faster simulations. Third, we do not need to split up the electron microfield into its low-frequency and high-frequency component, as done in classical MD simulations. Equation~\ref{eq:microfield_equation} gives the low-frequency microfield directly.

\subsection{Line Shape Calculations}

We calculate line shape profiles with our new microfields to test their impact on plasma diagnostics. We use the \textsc{Balrog} code \citep{Gomez21} to calculate all the line shape profiles presented in this paper. \textsc{Balrog} is a semi-analytic all-order full-Coulomb quantum model. It solves for the electron broadening, and includes the ion broadening using a plasma microfield. This microfield is incorporated following Eq. \ref{eq:fund_ls_eq}, as is done in other semi-analytic line broadening approaches. Any microfield can be used; however the standard choice in the \textsc{Balrog} code is to use the APEX model \citep{Iglesias2000}. 

\section{Results} \label{sec:results}
Here we present plasma microfields and line shape profiles from configuration-resolved pseudoatom molecular dynamics simulations at a range of conditions relevant to recent laboratory experiments. Among the selected conditions is a solid density Mg plasma, comparable to experiments preformed at the LCLS measuring K-$\alpha$ emission in solid density plasmas \cite{Ciricosta2012,Ciricosta16}. We also consider iron (Fe) at approximately the solar interior conditions probed by the Fe opacity experiments at the Sandia Z-Machine \cite{Bailey15,Nagayama19}. Finally we evaluate an oxygen (O) plasma at similar temperature and density to the O opacity experiments preformed at the Z-Machine and the National Ignition Facility (NIF) \cite{Mayes23}.

\subsection{Comparison Microfields}

We compare PAMD to two different microfield models. 
One is the APEX code \citep{Iglesias1985,Iglesias2000}, which uses the adjustable parameter exponential approximation and is commonly used in analytic line shape codes.
The second is the Potekhin Monte Carlo model \citep{Potekhin2002}, which is expected to be valid for Coulomb coupling parameters $\Gamma \leq 100$. Specifically, we use the fitting function for the microfield at a plasma ion provided in the appendix of \citet{Potekhin2002}.

For each comparison the ``APEX" microfield is calculated using the radiator/perturber concentration, radiator/perturber $\bar Z$, free electron density, and temperature as specified in or output by PAMD. We include APEX degeneracy corrections. The ``Potekhin" microfields are calculated using the coupling parameter and screening parameter determined from the PAMD results. For clarity and ease of comparison, all microfields are given in atomic units rather than a dimensionless $\beta$ parameter as is often done.

Each model makes different assumptions about the radiator. The APEX model assumes a two species plasma with separate net charges for the radiator and perturber species. The Potekhin model assumes a single species plasma where the perturbers and radiator share the same net charge. The Potekhin model is therefore only an appropriate comparison when the radiator net charge is close to the plasma net charge. It is therefore only included in figures \ref{fig:Mg_mf_avg}, \ref{fig:fe_mf},\ref{fig:fe_mf_log}, and \ref{fig:O_mf_lin}. 

\subsection{Solid Density Magnesium}

\begin{figure}
    \centering
    \includegraphics[width=0.9\columnwidth]{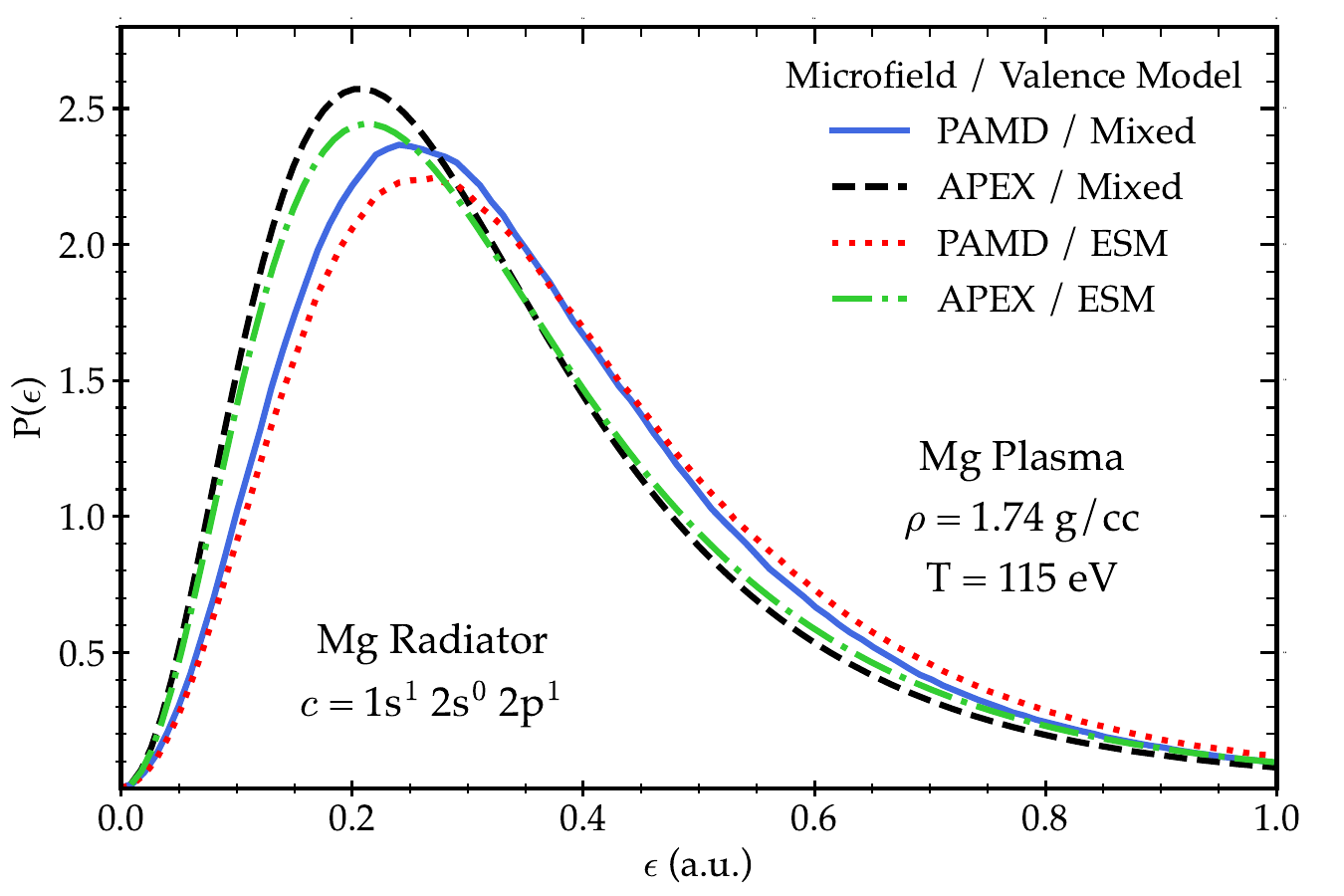}
    \caption{PAMD and APEX Mg microfields in a solid density Mg plasma. Two treatments of the valence shell electrons, discussed in section \ref{sec:valence_shell}, are compared. The mixture configuration is $c_{mix}$ = 1s$^1$ 2s$^0$ 2p$^1$ 3s$^0$ 3p$^0$ 3d$^0$ + FD, and the excited state configuration is $c_{ESM}$ = 1s$^1$ 2s$^0$ 2p$^1$ + FD.}
    \label{fig:Mg_mf}
\end{figure}

In figure \ref{fig:Mg_mf}, we compare PAMD microfields to APEX microfields for a solid density Mg plasma at a temperature of 115~eV. The radiator is in an excited Mg configuration corresponding to the initial state of a K-$\alpha$ emission transition at approximately the conditions measured experimentally in \citet{Ciricosta2012,Ciricosta16}, which has recently been used to test continuum lowering prescriptions \citep{Thelen24,PerezCallejo24}.

As in \citet{Thelen24}, we treat the radiator valence shell with the ESM approach. For comparison purposes we also include microfields calculated with the mixing approach. The appropriate APEX comparison microfield depends on the valence model because our definition of $\bar Z$ depends on the valence model. At these conditions, for a 1s$^1$ 2s$^0$ 2p$^1$ radiator, PAMD gives $\bar Z_{mix} = 10$ and $\bar Z_{ESM} \approx 9$.

\begin{figure}
    \centering
    \includegraphics[width=0.9\columnwidth]{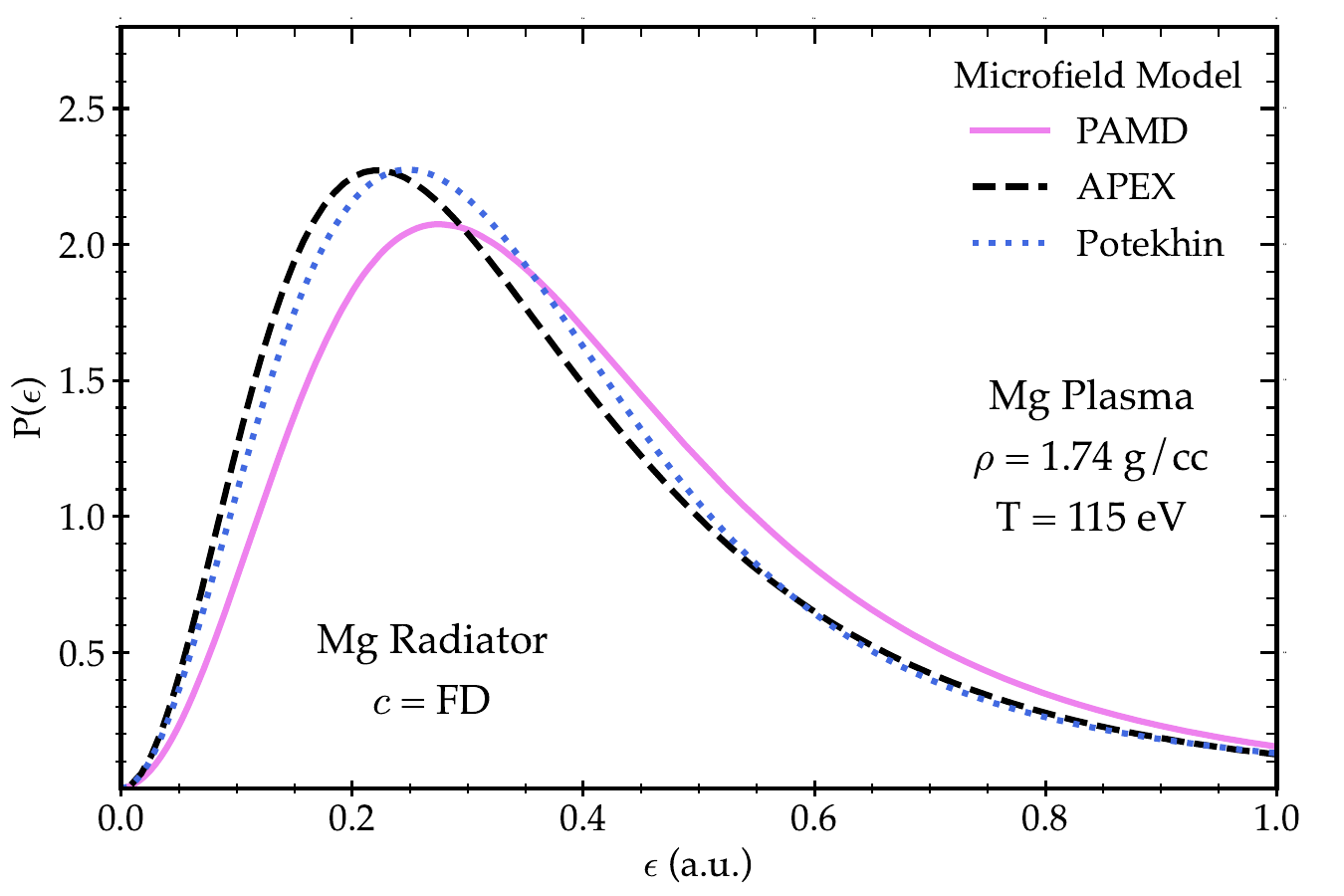}
    \caption{PAMD and APEX Mg microfields in a solid density Mg plasma, for a FD-configuration radiator.}
    \label{fig:Mg_mf_avg}
\end{figure}

Normally a 1s$^1$ 2s$^0$ 2p$^1$ configuration could be referred to as ``He-like" because it has 2 bound electrons. However, this naming convention is misleading when some bound states are treated as hybridized or as occasionally pressure ionized. In our 1s$^1$ 2s$^0$ 2p$^1$ + FD configuration there are still only 2 \textit{tightly} bound electrons. But as our model allows for fractional contributions from \textit{loosely} bound electrons, we attempt to avoid classifying atoms according to their charge state when possible. The quantity $\bar Z$ is misleading for similar reasons. Reducing our reliance on $\bar Z$ is a goal of future work.

The discrepancy between PAMD and APEX at these plasma conditions, seen in figure \ref{fig:Mg_mf}, is not unique to this specific configuration. In figure \ref{fig:Mg_mf_avg}, we compare microfields in the same plasma for an average configuration radiator and find similar levels of disagreement. We therefore believe that the source of this disagreement is unlikely to be the valence shell treatment or the configuration resolution. Instead, we expect that including N-body effects and our improved treatment of screening are the most likely causes. This issue is discussed in greater detail in section \ref{sec:model_comp}.

\begin{figure*}
    \centering
    \includegraphics[width=2\columnwidth]{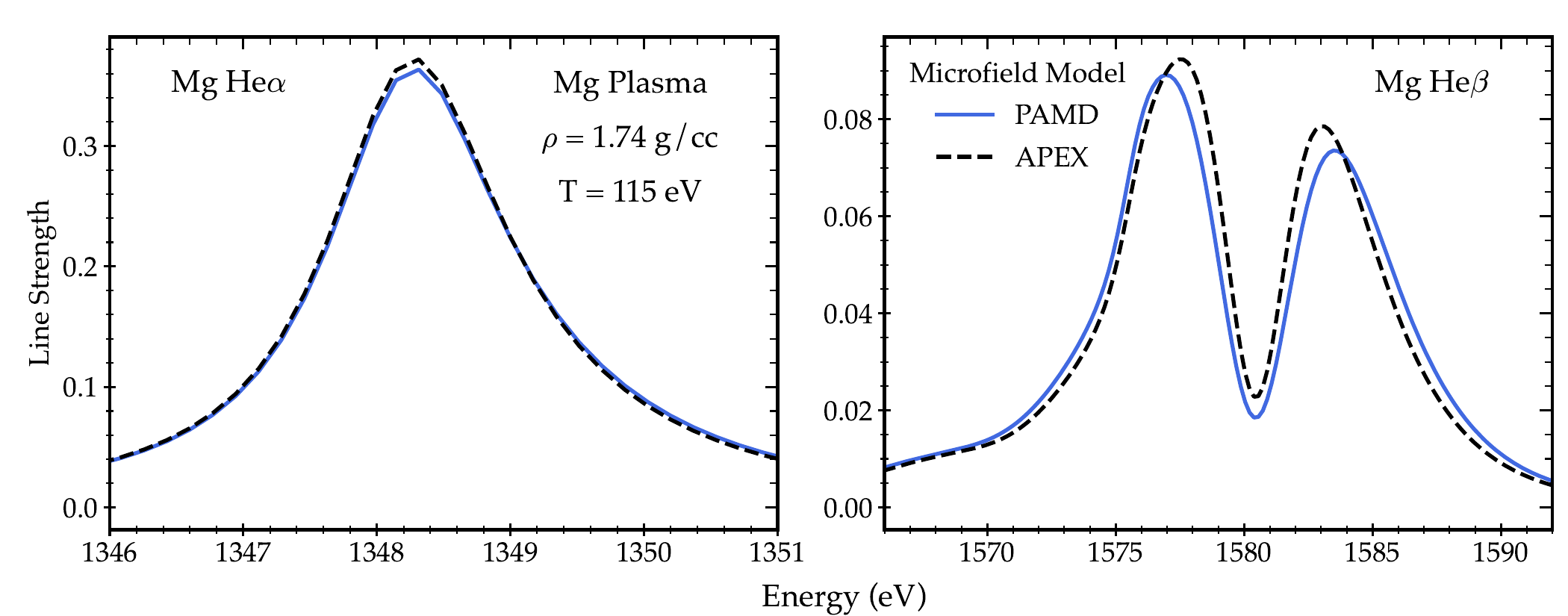}
    \caption{He$\alpha$ and He$\beta$ line shape profiles calculated using excited state model PAMD and APEX microfields. }
    \label{fig:Mg_Hea_Heb}
\end{figure*}

In figure \ref{fig:Mg_Hea_Heb}, we present Mg He$\alpha$ and Mg He$\beta$ absorption lines. We compare absorption profiles so that the same set of microfields can be used for both lines. He$\alpha$ is a relatively isolated line, so the ion broadening is dampened \citep{Alexiou14}, and the deviations are not significant enough to cause substantial changes in the line broadening. He$\beta$, by comparison, is less isolated and therefore more sensitive to changes in the microfield, leading to an increase in the FWHM of $\tildetext8\%$ when the PAMD results are used.

\subsection{Pulsed Power Iron} \label{sec:iron_results}

\begin{figure}
    \centering
    \includegraphics[width=0.9\columnwidth]{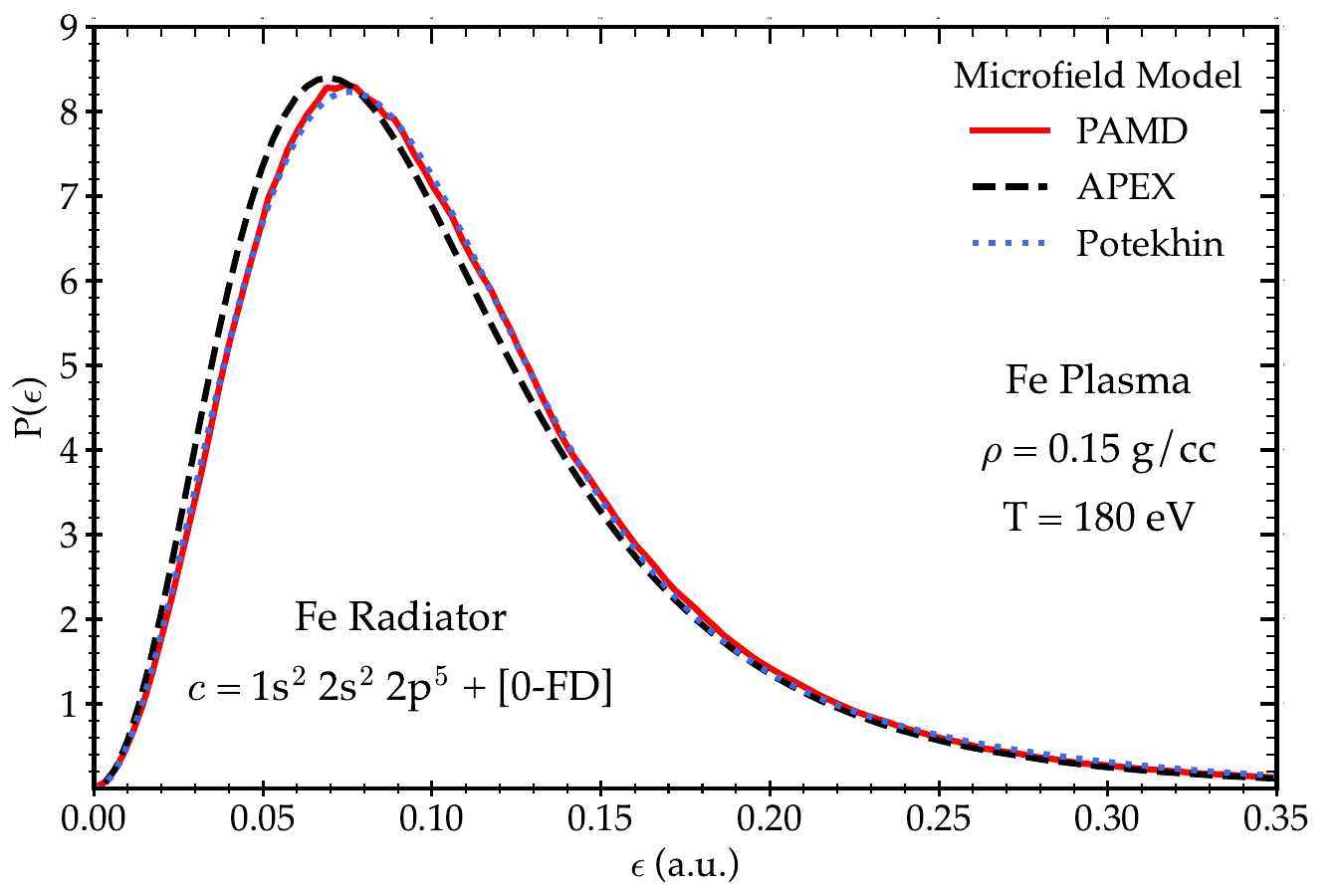}
    \caption{PAMD, APEX, and Potekhin microfield for an iron radiator in an iron plasma. }
    \label{fig:fe_mf}
\end{figure}

Figure \ref{fig:fe_mf} compares the pseudoatom microfield to the APEX and Potekhin microfields for an Fe radiator in a dense Fe plasma with 9 tightly bound electrons, using the mixing model. These conditions correspond to a plasma that is less strongly coupled than in the previous Mg case. The agreement between models is excellent, especially between PAMD and Potekhin. The PAMD radiator $\bar Z_{rad} \approx $ 17.00 and perturber $\bar Z_{per} \approx $ 16.97 are in good agreement at these conditions.

\begin{figure}
    \centering
    \includegraphics[width=\columnwidth]{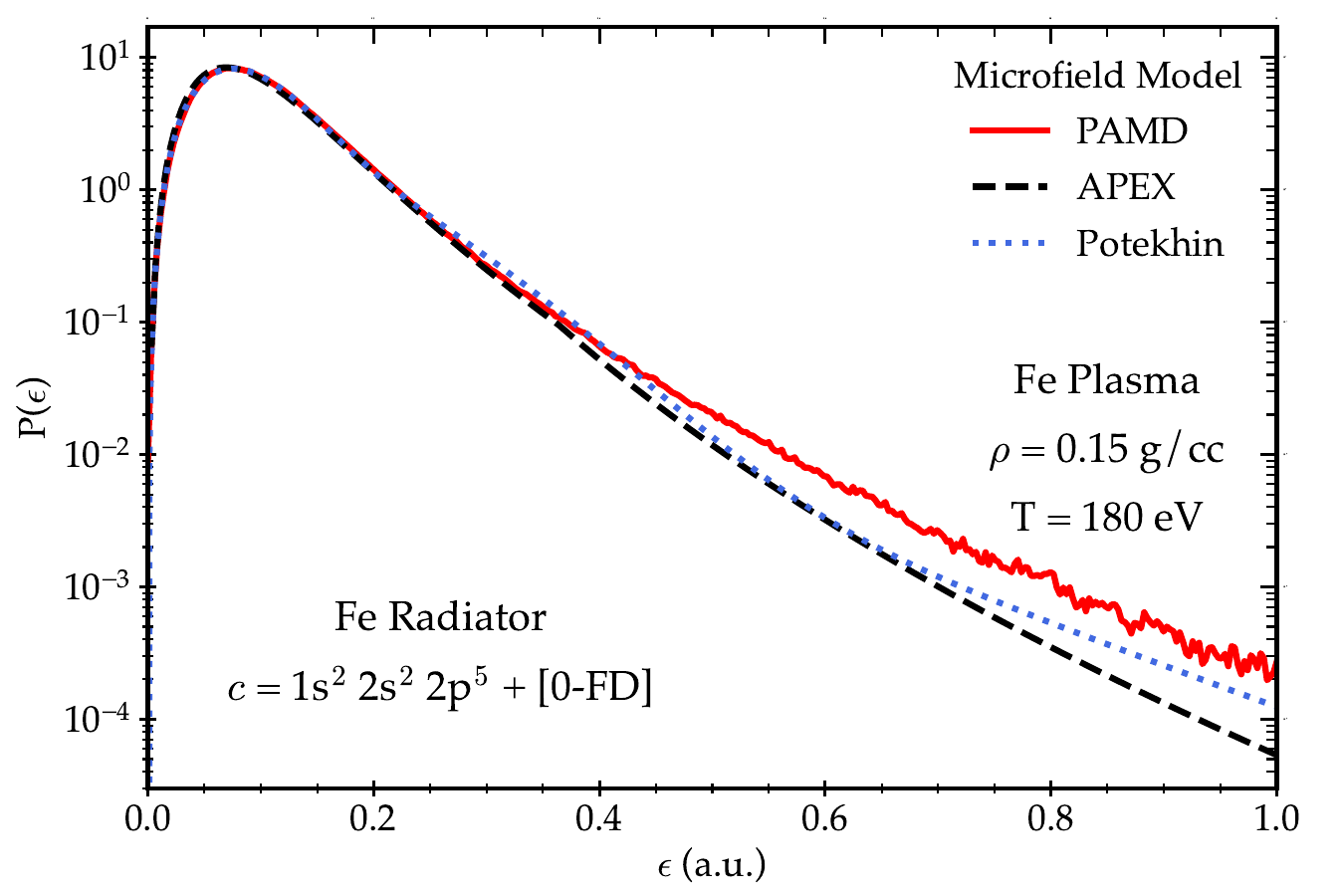}
    \caption{Same as figure \ref{fig:fe_mf}, but in log scale to highlight the high-$\epsilon$ tail.}
    \label{fig:fe_mf_log}
\end{figure}

Stronger microfield deviations are present in the high-$\epsilon$ tail, as shown in figure \ref{fig:fe_mf_log}. A stronger microfield in the high-$\epsilon$ regime generally implies a broader line shape, particularly in the wing. However, the far line wing is challenging to resolve experimentally when there are many lines in close proximity to each other, as is the case in the Fe opacity experiment mentioned previously.

Figure \ref{fig:fe_mf2} shows the same results as in figure \ref{fig:fe_mf}, but for a FeMg plasma. This choice is more comparable to the Fe opacity experiment, which uses a FeMg sample \citep{Bailey15}. Here the radiator and perturber average $\bar Z$ values differ substantially because there are both Fe and Mg perturbers, so the Potekhin model is not compared. The level of agreement between APEX and PAMD is similar for both plasma compositions.

\subsection{Solar Interior Oxygen}

\begin{figure}
    \centering
    \includegraphics[width=\columnwidth]{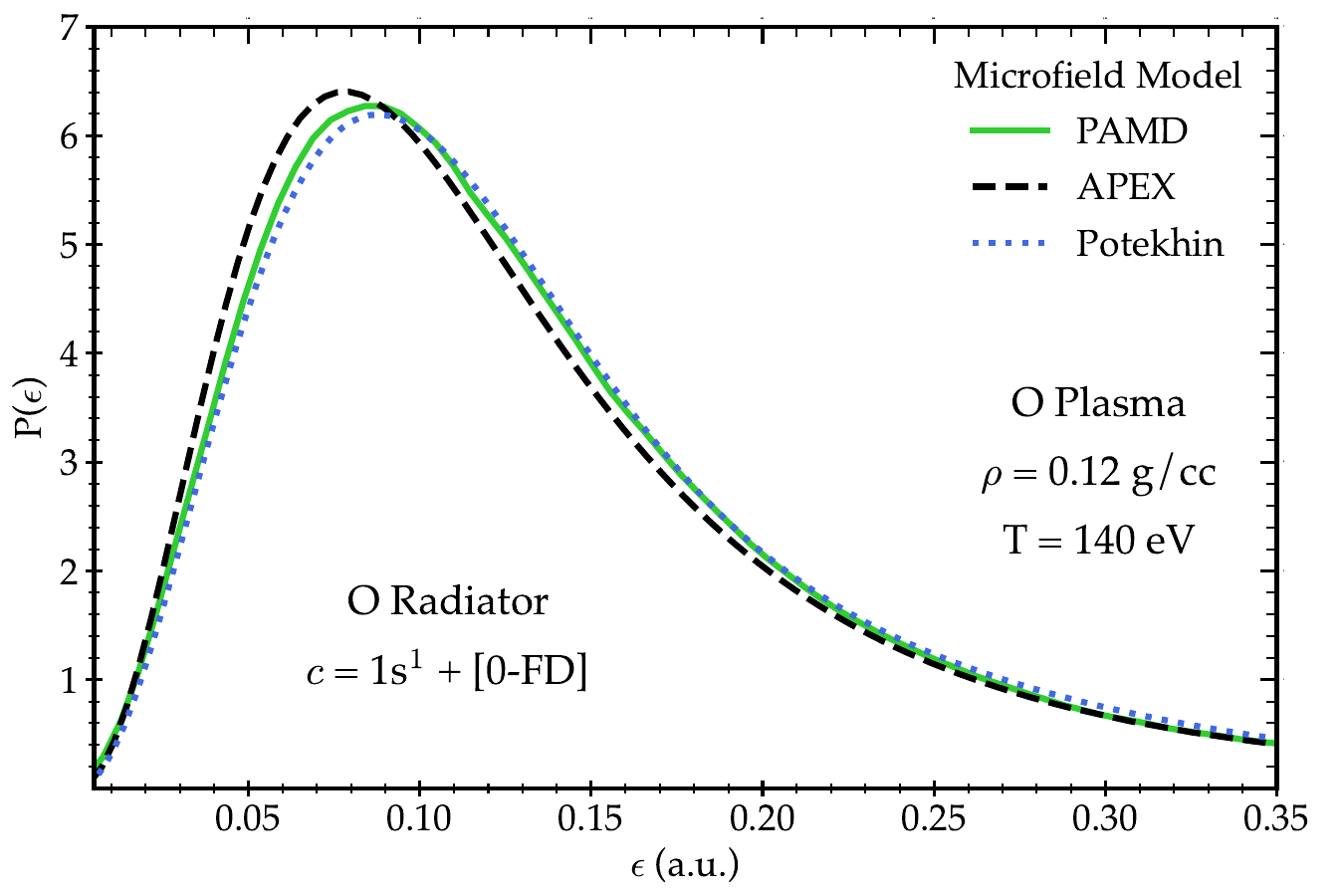}
    \caption{Microfield comparison for a hydrogen-like oxygen atom in an oxygen plasma with density of 0.12 g/cc and a temperature of 140 eV.  }
    \label{fig:O_mf_lin}
\end{figure}

Figure \ref{fig:O_mf_lin} shows the microfield comparison for an O radiator in a dense O plasma. The temperature and density regime is similar to the solar convection zone boundary and the O opacity experiments preformed at the Sandia Z-Machine and NIF \cite{Mayes23}. The PAMD results are again in good agreement with the APEX and Potekhin model microfields, as in the Fe case explored previously. The radiator $\bar Z_{rad} \approx 7$ and perturber $\bar Z_{per} \approx 6.92$ are also in decent agreement at these conditions.

\begin{figure}
    \centering
    \includegraphics[width=\columnwidth]{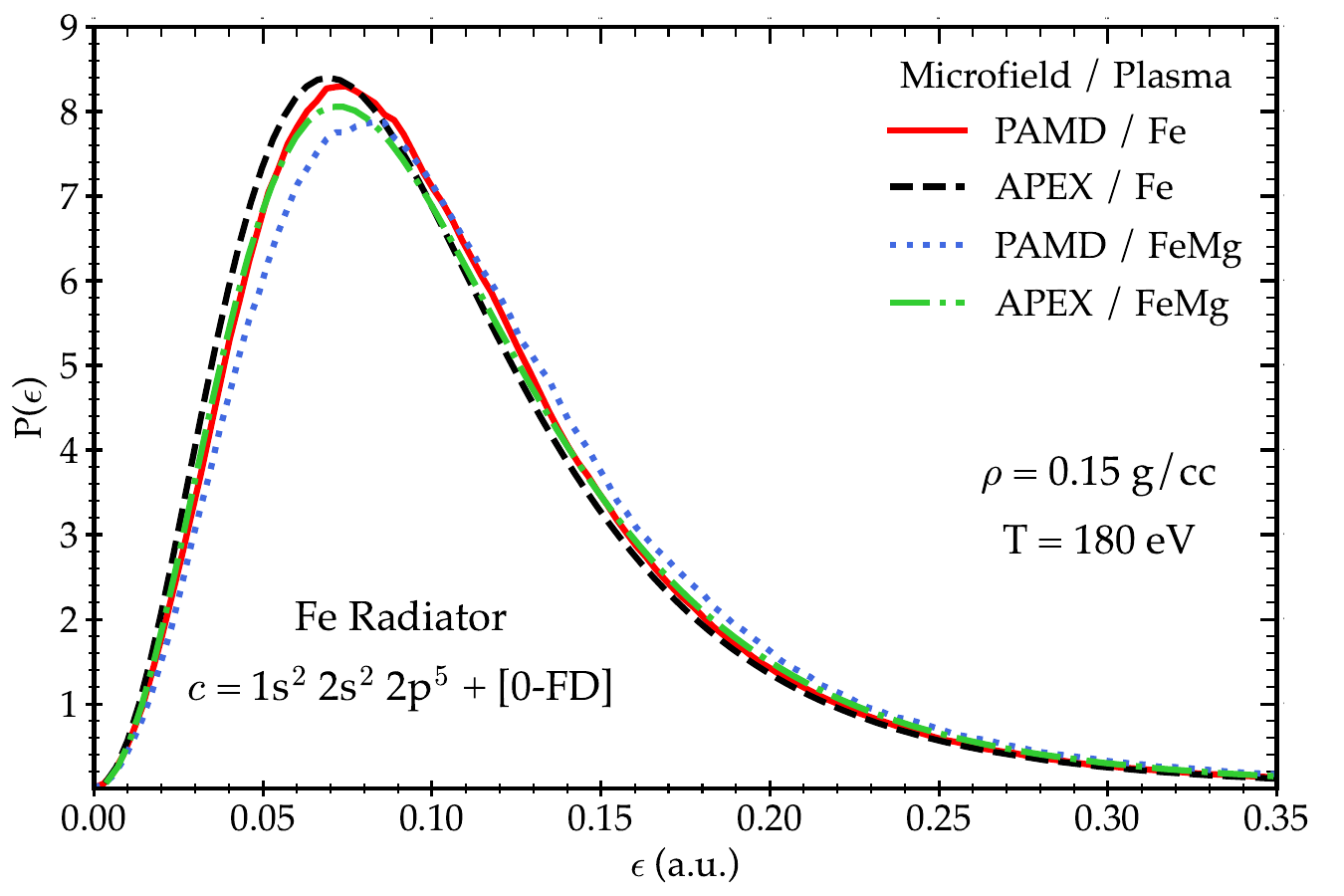}
    \caption{Same as figure \ref{fig:fe_mf} but with two different plasma compositions considered, Fe alone and an FeMg mixture. }
    \label{fig:fe_mf2}
\end{figure}

\begin{figure}
    \centering
    \includegraphics[width=\columnwidth]{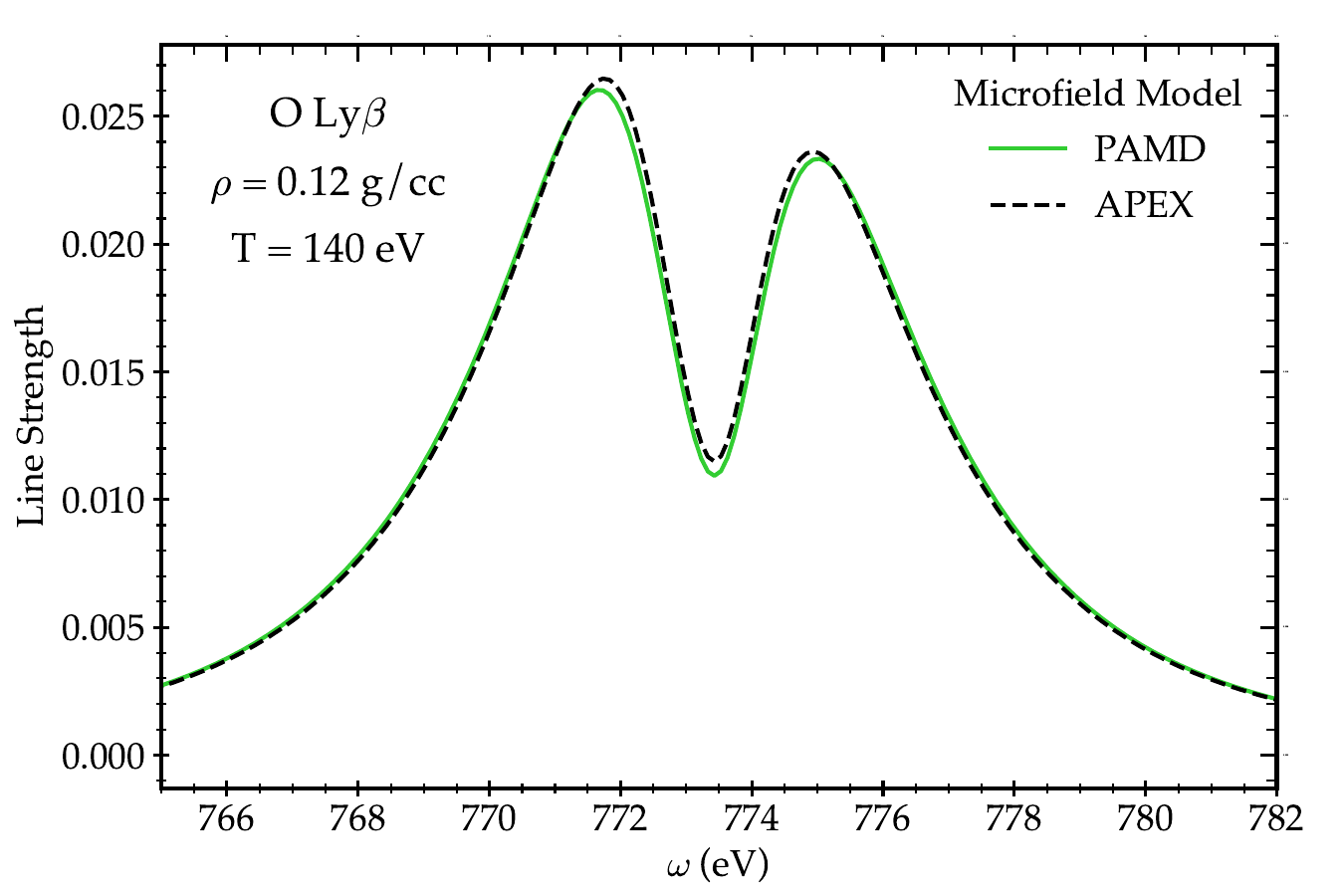}
    \caption{Oxygen Ly$\beta$ line shape calculated with the PAMD and APEX microfields in figure \ref{fig:O_mf_lin}. The FWHM differs by $ 3 \% $.}
    \label{fig:O_LyB_LS}
\end{figure}

In figure \ref{fig:O_LyB_LS}, we present an oxygen Ly$\beta$ line shape calculated with the different microfields from figure \ref{fig:O_mf_lin}. As expected, the slight shift towards stronger electric fields in the PAMD microfield compared to APEX causes a slight increase in broadening. The difference, however, is largely negligible. It would be challenging to differentiate between the two in a laboratory experiment.

We note that the plasma composition is different here than in the laboratory or in the solar interior; O is primarily perturbed by H near the solar convection zone boundary, while a number of different perturbers are present in recent laboratory experiments including O, Mg, and Si. There could be more substantial differences depending on the perturber composition. However, the consistency found in figure \ref{fig:fe_mf2} suggests that this is unlikely. 

 \setlength{\tabcolsep}{6pt} 
\renewcommand{\arraystretch}{1.2} 
\begin{table*}
\centering
\begin{tabular}{ | c | c || c | c | c | c | }
 \hline
 & Composition & Mg  & O & FeMg \\
 Plasma & $T$ (eV)  & 115  & 140  & 180 \\
 Conditions & $ \rho$ (g/cc) & $1.74$ &  $0.12$   & $0.15$  \\
 & $\Gamma_i $ & 4.3 & 1.3 & 3.0 \\
  & $\omega_i^{-1}$ (fs) &  $ 2.3 $ & $6.5$ & $7.6$ \\
 
\hline
 
Radiator & Configuration & Mg 1s$^1$ 2s$^0$ 2p$^1$ & O 1s$^1$ & Fe 1s$^2$ 2s$^2$ 2p$^5$      \\

 $ $ & Lifetime (fs) & $ 0.48$& $23$ & $15$ \\
 \hline
\end{tabular}
\caption{Conditions and characteristic timescales for selected magnesium, oxygen, and iron-magnesium cases. The ion coupling parameter is defined as $\Gamma_i = \bar Z ^2 / r_{s} T$, where $r_s$ is the Wigner-Seitz radius. The inverse ion plasma frequency is given as $\omega_i^{-1} = (4\pi n_i \bar Z^2 / m_i)^{-1/2}$. The configuration lifetimes are generated with the Los Alamos suite of atomic physics codes~\citep{Fontes2015}.}
\label{tab:timescale_tab}
\end{table*}

\section{Discussion} \label{sec:discussion} 

We have presented a novel approach to calculating low-frequency microfields relevant for spectral line diagnostics using configuration-resolved pseudoatom molecular dynamics. Here we discuss some limitations to the presented approach, possible future improvements to this technique, and its current level of success.

\subsection{Averaged Perturbers}
 A deficiency of our method is that configuration-resolved pseudoatoms are used only for the plasma radiator. Plasma perturbers are only included as FD-occupied pseudoatoms. A more robust approach would be to populate the simulation exclusively with configuration-resolved pseudoatoms where the populations are determined from the charge state distribution and configuration probabilities \citep{White2022,Starrett23}. However, such an approach is computationally challenging due to the large number of configurations involved. We therefore do not attempt such a mixture here.

\subsection{Time Dependence}
The microfields calculated in this paper, as well as in     the models we compare against, partially neglect the time dependence of the problem by assuming the radiator has not recently undergone a change in configuration. Physically, we should not expect the particle distribution to immediately adjust after an excitation, decay, ionization, or recombination occurs. The average microfield should instead evolve over the course of some characteristic `relaxation' timescale as the ion distribution approaches a steady-state solution. We expect that this time dependence can safely be neglected if the relaxation time is short compared to the lifetime of the radiator. 

Table \ref{tab:timescale_tab} presents inverse ion plasma frequencies and configuration lifetimes for the different pseudoatom species and plasma conditions explored in section \ref{sec:results}.
We take the inverse ion plasma frequency as a characteristic relaxation timescale. This choice is physically motivated~\citep{HauRiege15,HauRiege17} but not rigorous and is intended only as an approximation. 
We estimate the lifetimes of each configuration from atomic rates generated with the Los Alamos suite of atomic physics codes~\citep{Fontes2015}. 

For the two ground-configuration cases, the lifetime is a factor of a few greater than the relaxation timescale. For the excited Mg configuration, the lifetime is a few times smaller than the relaxation timescale. Broadly speaking, the O and Fe microfields presented here are therefore expected to be less sensitive to the configuration time history of the radiator, while the Mg microfields are likely to be more sensitive. However, the relatively small (less than an order of magnitude) difference in timescales for each case and approximate nature of the relaxation timescale make it challenging to definitively assert to what degree the time history will influence the microfields. We therefore conclude that a more detailed analysis on this topic is warranted in the future.

\subsection{Convergence}

\begin{figure}
    \centering
    \includegraphics[width=\columnwidth]{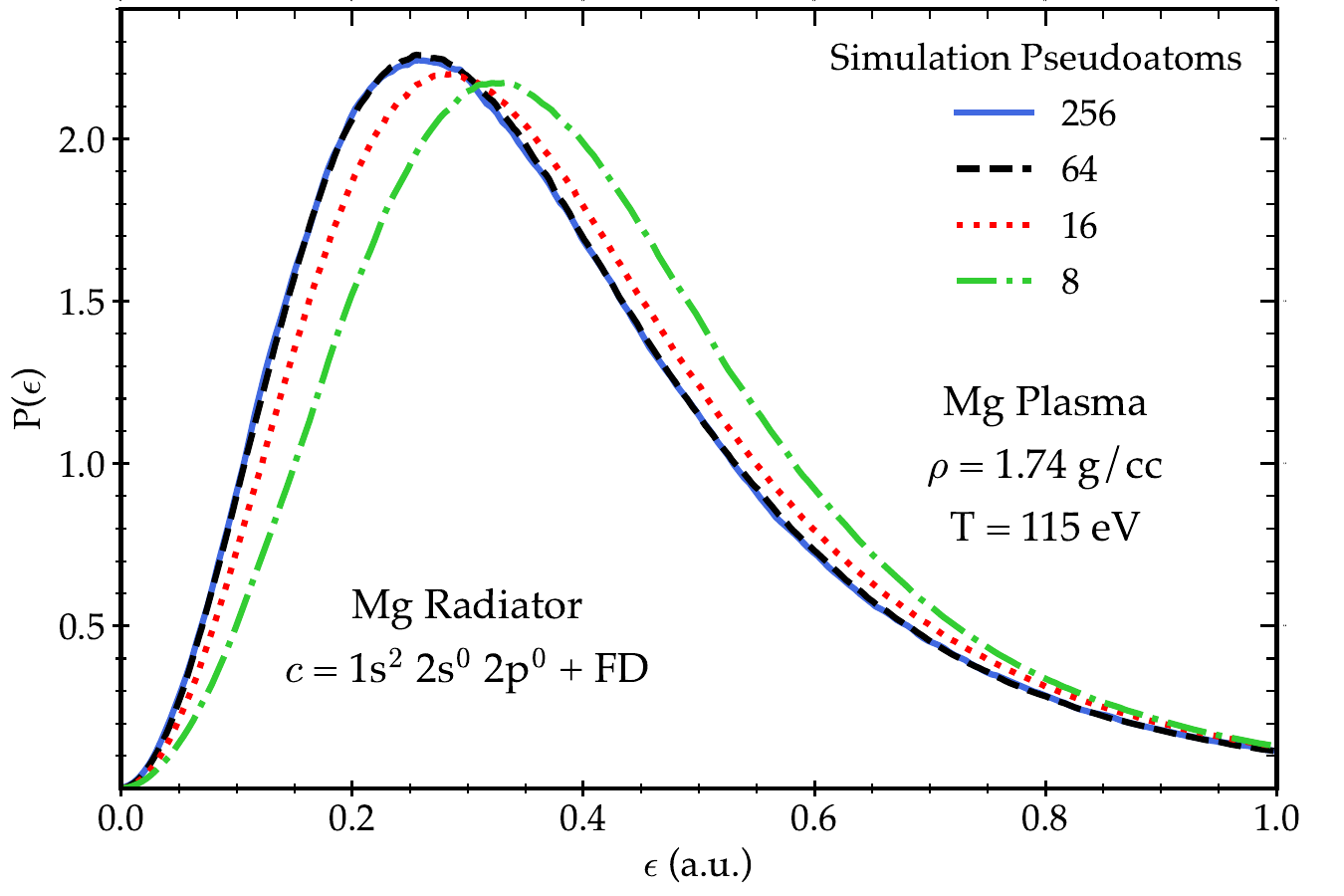}
    \caption{Microfield distributions calculated with a varying number of pseudoatoms.}
    \label{fig:Mg_convg}
\end{figure}

We note that the simulation and microfield results are sensitive to the total number of pseudoatoms included. We check for convergence by increasing the size of the simulation until the results do not vary appreciably. Figure \ref{fig:Mg_convg} demonstrates this process for a Mg microfield in a solid density Mg plasma. We typically find that $\gtrsim$100 total pseudoatoms is sufficient to achieve a well converged result, as is the case in figure \ref{fig:Mg_convg}.

\subsection{Model Comparisons} \label{sec:model_comp}
The convergence of the PAMD microfield to both the APEX and Potekhin models in the lower density plasmas explored here bolsters our confidence in the validity of our model. In relatively low density plasmas, a screening length should be adequate to account for particle correlation effects, so we expect to agree well with other microfield models. Strong agreement with the APEX and Potekhin models in figures \ref{fig:fe_mf} and \ref{fig:O_mf_lin} is therefore encouraging.

However, there is a more substantial difference between the solid density Mg microfields in figure~\ref{fig:Mg_mf}. We principally attribute this difference to the fact that we are preforming an N-body simulation and modelling ion-electron correlations (Eq. \ref{eq:potential_eq}) using linear response theory instead of a Yukawa-type screening prescription. We expect our pseudoatom approach to preform better at high-density conditions where the validity of simple screening prescriptions starts to break down. Previous comparisons between APEX and MD simulations \citep{Demura09,Iglesias2000} have found similar levels of disagreement at high coupling conditions, especially in the high-$\epsilon$ tail of the microfield.

Future comparison against other MD microfield codes \citep{HauRiege17,Calisti24} would be beneficial. The fundamentally different treatments of free electrons will, however, make direct comparisons challenging, as there is some freedom in determining the appropriate averaging interval for calculating the low-frequency microfield when individual ions and free electrons are resolved.

\section{Conclusions}\label{sec:conclusion}
We have presented a new approach to calculating plasma microfields using pseudoatom molecular dynamics (PAMD) simulations. This capability is enabled with the introduction of configuration-resolved pseudoatoms, with user-specified integer bound state occupations, to PAMD for the first time.

We compare two different methods of defining electron configurations in a consistent manner across pressure ionization boundaries. Both are approximate solutions, and should be improved upon in the future, but we find that they can be effective at ensuring a computationally smooth transition across densities. 

Using configuration-resolved PAMD, we calculate microfields relevant to a number of recent warm and hot dense matter laboratory experiments at facilities including the Linac Coherent Light Source (LCLS) \citep{Ciricosta16}, Sandia Z-Machine \citep{Bailey15}, and the National Ignition Facility (NIF) \citep{Mayes23}. Comparisons to established microfield codes show good agreement at the lower density conditions that we explore, where analytic models are expected to be more reliable. Stronger deviations are found in the case of solid density Mg, where our improved treatment of screening and N-body effects is expected to be more impactful.

Finally we present spectral line shapes calculated with our new microfields. Small changes to the broadening, on the order of a few \%, are found when compared to other microfield calculations. The differences are generally small enough that they would be challenging to detect in experimental or astrophysical data.

\section*{CRediT authorship contribution statement}

\textbf{J.R. White:} Investigation, Writing – original draft, Writing – review \& editing, Formal Analysis, Visualization, Methodology. \textbf{C. J. Fontes:} Investigation, Writing – review \& editing, Supervision. \textbf{M. C. Zammit:} Investigation, Writing – review \& editing, Supervision. \textbf{T. A. Gomez:} Investigation, Writing – review \& editing, Formal Analysis. \textbf{C. E. Starrett:} Investigation, Writing – review \& editing, Supervision, Methodology.

\section*{Declaration of competing interest}

The authors declare no conflicts of interest.

\section*{Data Availability}

The data will be made available upon request.

\section*{Acknowledgements}
J.W. acknowledges support from the United States Department of Energy National Nuclear Security Administration SSGF program under DE-NA0003960, and the Wootton Center for Astrophysical Plasma Properties
under U.S. Department of Energy Cooperative Agreement No. DE-NA0004149. J.W., C.F., M.Z., and C.S. acknowledge support from the Los Alamos National Laboratory (LANL) ASC PEM Atomic Physics Project. LANL is operated by Triad National Security, LLC, for the National Nuclear Security Administration of the U.S. Department of Energy under Contract
No. 89233218NCA000001.

\clearpage


\bibliographystyle{elsarticle-num-names} 
\bibliography{main}

\end{document}